\documentclass[aps,twocolumn,floatfix,nofootinbib,pra,longbibliography]{revtex4-1}
\usepackage[utf8]{inputenc}
\usepackage{lipsum}
\usepackage{graphicx}
\usepackage{amsmath,amssymb}
\usepackage{braket}
\usepackage{mathtools}
\usepackage{hyperref}
\usepackage{multirow}
\usepackage{color}
\usepackage[normalem]{ulem}
\usepackage{amsfonts}
\usepackage{bbm}

\newcommand{\bea}{\begin{eqnarray}}
\newcommand{\eea}{\end{eqnarray}}
\newcommand{\beq}{\begin{equation}}
\newcommand{\eeq}{\end{equation}}

\def\ket#1{{|#1\rangle}}
\def\bra#1{{\langle #1 |}}

\begin{document}

\title{Push-down automata as sequential generators of highly entangled states}

\author{Sarang Gopalakrishnan}
\affiliation{Department of Electrical and Computer Engineering, Princeton University, Princeton, NJ 08540}

\begin{abstract}

We exploit the duality between quantum channels and sequentially generated states to construct families of highly entangled states that undergo phase transitions. These highly entangled states can be sequentially generated by collecting the emitted radiation from an open quantum system. In the dual perspective, the open system is regarded as a quantum-state-generating machine. A nontrivial class of such machines are quantum push-down automata, which can be used to create highly entangled states including the ground state of the spin-2 Motzkin spin chain parametrically faster than adiabatic evolution. We construct generalizations of the Motzkin state that can be efficiently sequentially generated. 

\end{abstract}
\maketitle

%


Most quantum states are hard to prepare~\cite{PhysRevD.97.086015}, but also useless. The atypical states we know how to \emph{use} are often specified, perhaps implicitly, in terms of recipes for making them. 
%
%
 Of these, the best-studied class consists of ground states of local Hamiltonians, which naturally occur in low-temperature matter, or can be prepared through adiabatic evolution from a reference state. 
 Programmable quantum systems allow for more recipes and more states: e.g., those reachable by applying quantum circuits to a reference state, or by operating on a lightly entangled state using measurement and feedback~\cite{PhysRevLett.127.220503, zhu2022nishimori, foss2023experimental, iqbal2023topological} (as in measurement-based quantum computation~\cite{briegel2009measurement, PhysRevA.76.052315}).
Ground states have been classified into distinct phases of matter; for the states that result from more general preparation protocols, the analogous questions have been less studied.
 
Here we explore one such protocol, the sequential generation of quantum states~\cite{PhysRevLett.95.110503, PhysRevA.77.052306, PhysRevLett.128.010607, PhysRevLett.105.260401, PhysRevA.95.032312, astrakhantsev2022time, barratt2021parallel, PhysRevResearch.3.033002, PhysRevLett.128.150504, anand2022holographic}. This protocol underlies photonic quantum computing~\cite{PhysRevLett.103.113602, PhysRevLett.105.093601, PhysRevLett.95.010501, pichler2017universal}, and can be specified as follows. We consider a Markovian open quantum system called an ``emitter,'' which continually leaks information into its environment. Radiation emitted at different times is entangled through its interaction with the emitter; we are interested in characterizing the state of this emitted radiation. When the emitter has a finite Hilbert space dimension $\chi$, the emitted state is a matrix-product state (MPS) of bond dimension $\chi$. A finite-size open system generically does not undergo phase transitions, so MPSs at finite $\chi$ are too restrictive an ansatz to capture phase transitions except in special cases~\cite{PhysRevResearch.4.L022020}. Instead, one needs an emitter with an infinite-dimensional Hilbert space. Although some specific cases of infinite emitters~\cite{PhysRevLett.105.093601, pichler2017universal} have been explored, little is known about such emitters in general. (One exception is the pioneering recent works~\cite{PhysRevX.12.011045, PRXQuantum.2.040319}.)

We consider infinite emitters that can be tuned across nonequilibrium phase transitions, viewed as sequential state-generating machines. We find conditions under which such machines efficiently and reliably emit highly entangled states. We introduce this idea by presenting a protocol to construct the spin-2 Motzkin state~\cite{PhysRevLett.109.207202, movassagh2016supercritical, zhang2017novel, alexander2021exact} (in which contiguous subregions of size $\ell$ have entanglement scaling as $\sqrt{\ell}$). This protocol is parametrically faster than adiabatic evolution, and also outperforms classical MPS simulations. 
We generalize this construction to a larger family of machines, which are closely related to push-down automata~\cite{sipser2012introduction, moore2000quantum, qiu2002quantum}. Our approach yields two efficiently constructable generalizations of the Motzkin state with entanglement that scales as $\ell^{3/4}$ and $\ell/O(\log \ell)$ respectively.

\begin{figure}[b]
\begin{center}
\includegraphics[width=0.4\textwidth]{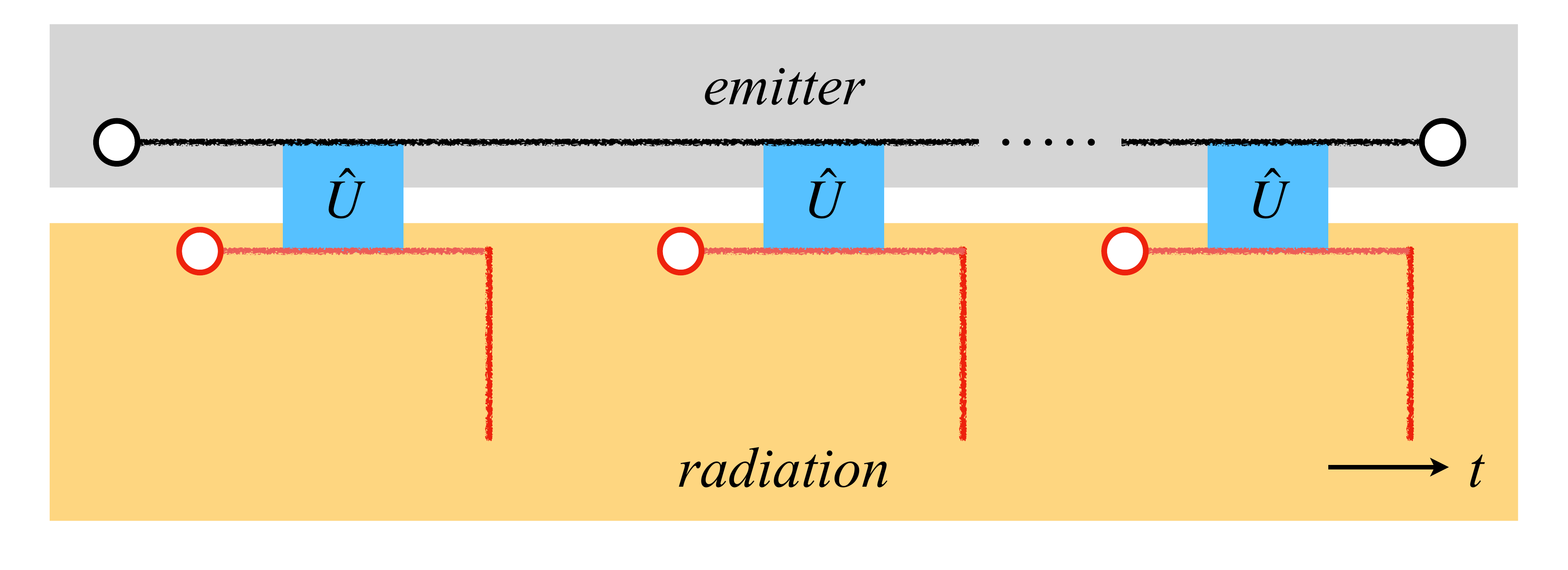}
\caption{State-generating (``dual'') perspective on an iterated quantum channel: the reservoir repeatedly sends qubits, initialized in $\ket{0}$, into the emitter, and the radiation that comes out is harvested to make a state in $\mathcal{H}_{\mathrm{rad}} \equiv \bigotimes_{n} \mathcal{H}_n$.}
\label{fig1}
\end{center}
\end{figure}

\emph{Setup}.---The emitter is a quantum system with Hilbert space $\mathcal{H}_{\mathrm{em}}$ that evolves according to a repeated quantum channel $\mathcal{E}$. The system is initialized in a reference state $\ket{\Phi}\bra{\Phi}$, which we take to be a pure state. In Stinespring form, the $n$th application of the channel involves bringing in a reservoir qubit, with Hilbert space $\mathcal{H}_n$, initialized in a reference state $\ket{0}\bra{0}$. A unitary $U$ acts on the emitter and the $n$th reservoir qubit, after which the $n$th reservoir qubit is radiated and no longer interacts with the emitter. The entire process is sketched in Fig.~\ref{fig1}. In the standard perspective one traces over the reservoir qubits and derives a quantum channel acting on the operators in $\mathcal{H}_{\mathrm{em}}$. In the \emph{dual} perspective we will consider here, one instead harvests the radiated qubits, which are entangled with one another and, in general, with the emitter. To terminate the process (after $N$ radiated qubits have been harvested) one disentangles the emitter from the radiation by performing a rank-1 measurement on it. Each measurement outcome generates a different entangled state on $\bigotimes_{n=1}^N \mathcal{H}_n$. To generate a \emph{particular} state one post-selects on the desired outcome. We will focus on machines that have no advance knowledge of $N$---they keep autonomously generating entangled qubits until the final measurement is performed. 

The case where $\mathcal{H}_{\mathrm{em}}$ is finite-dimensional has been studied in depth~\cite{PhysRevLett.95.110503, PhysRevA.77.052306}. This construction generates a canonical-form MPS with a bond dimension that is (at most) the dimension of $\mathcal{H}_{\mathrm{em}}$. In fact, the entanglement spectrum of the radiated state across a cut is closely related to the density matrix on $\mathcal{H}_{\mathrm{em}}$ at the time corresponding to that cut, as follows~\cite{PhysRevB.100.064309, PhysRevLett.123.210601}. The $n$th R\'enyi entropy $S_n$ between a subsystem of the first $\ell$ radiated qubits and the remaining $N - \ell$ qubits can be written entirely in terms of the channel acting on $\mathcal{H}_{\mathrm{em}}$, as $S_n = (1-n)^{-1} \log \mathrm{Tr}[(\rho_\ell O_{N - \ell})^n]$, where  $\rho_\ell = \mathcal{E}^\ell(\ket{0}\bra{0}), O_{N - \ell} = (\mathcal{E}^*)^{N - \ell}(\mathcal{E}^\ell(\ket{0}\bra{0})$, and $\mathcal{E}^*$ is the ``dual'' channel (i.e., the Heisenberg-picture version of $\mathcal{E}$). Assuming the channel has a unique steady state, for large enough $N - \ell$ any operator in $\mathcal{H}_{\mathrm{em}}$ gets projected onto the leading eigenvector of the dual channel, which is the identity matrix. In that case, $S_n$ is simply the $n$th R\'enyi entropy of the steady state of $\mathcal{E}$, i.e., the eigenvalue spectrum of the steady-state density matrix of the open system on $\mathcal{H}_{\mathrm{em}}$.

\emph{Warmup: spin-1 Motzkin chain}.---We now consider, as our simplest nontrivial example, a model in which the emitter is a single particle on a semi-infinite chain. We introduce this model from the standard perspective. The states $\ket{i}, i \geq 0$, span $\mathcal{H}_{\mathrm{em}}$; each can be interpreted as the sole particle being on site $i$. Hopping is fully incoherent, so the channel is specified by its action on populations $p_i \equiv \bra{i} \rho \ket{i}$, which is that of a classical Markov chain. The transition matrix is:
\beq
p_i (t + 1) = (1 - \gamma_{L,i} - \gamma_{R,i}) p_i(t) + \gamma_{L,i} p_{i+1}(t) + \gamma_{R,i} p_{i-1}(t).
\eeq
For $i \neq 0$ we choose site-independent rates $\gamma_L$ and $\gamma_R$; for $i = 0$ we set $\gamma_{R,0} = 0$ and $\gamma_{L,0} = \gamma_L$. The nature of the steady state of the channel under this Markov chain depends on the ratio of rates $\delta \equiv \gamma_R/\gamma_L - 1$. When $\delta < 0$, leftward hopping dominates and causes probability to pile up near the wall at $i = 0$. When $\delta > 0$, rightward hopping dominates and the probability moves ballistically to the right. At the critical point, $\delta = 0$, the dynamics is that of a random walk with a reflecting boundary, so a wavepacket initialized near the origin spreads out to a distance $\sim \sqrt{t}$ in time $t$, and its probability of reappearing at the origin decreases as $1/\sqrt{t}$. 
 
 \begin{figure}[t]
\begin{center}
\includegraphics[width=0.4\textwidth]{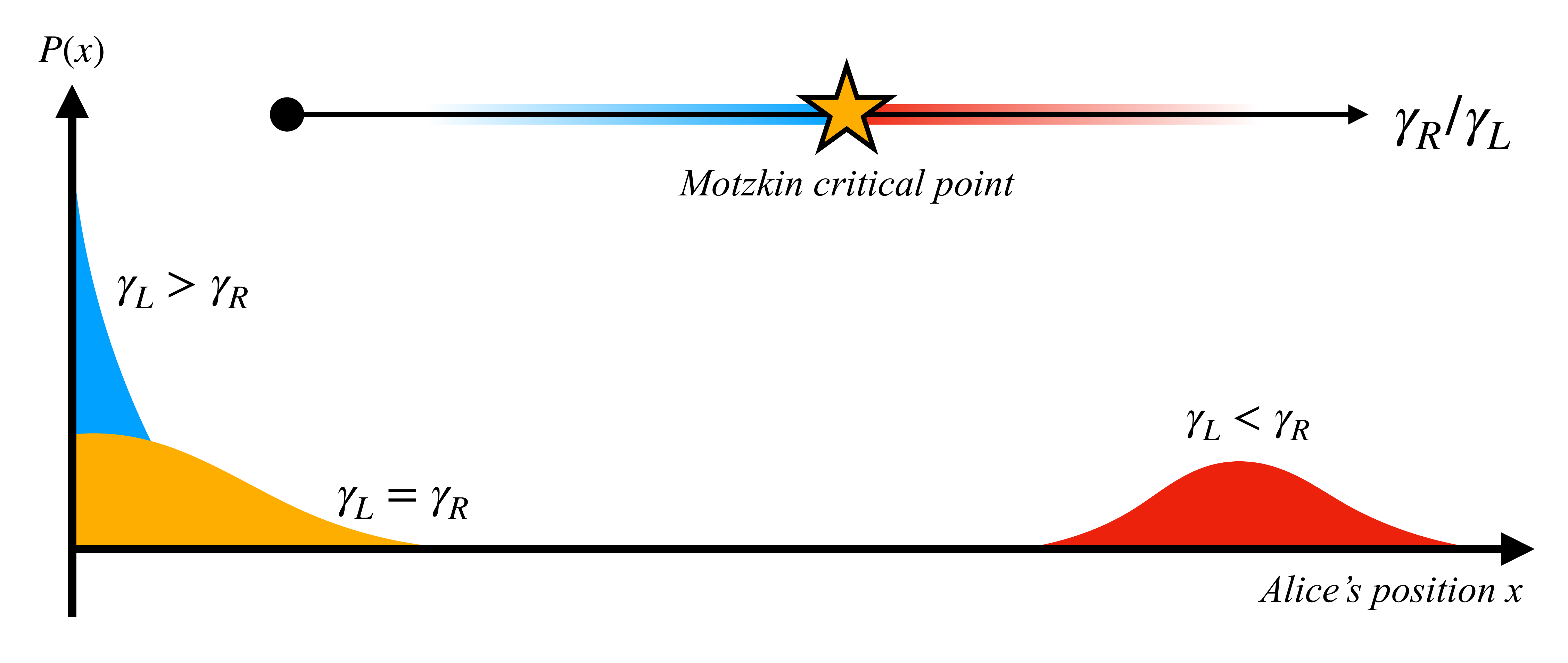}
\caption{Three phases of the biased random walk on the half-line with a reflecting boundary: pinned to the wall for $\gamma_L < \gamma_R$, escaping to infinity for $\gamma_L > \gamma_R$, and undergoing a random walk at the critical point (which corresponds, in the dual perspective, to the critical Motzkin state).}
\label{fig15}
\end{center}
\end{figure}
 
We now turn to the dual perspective. Since the particle (away from the origin) has three options at each time step, we need a channel with three Kraus operators. We dilate this by introducing reservoir degrees of freedom that are \emph{qutrits} with states $-1, 0, +1$. The reservoir is initialized in the state $\ket{0}$, and the system-reservoir unitary acts on this state as $\ket{n, 0} \mapsto \sqrt{1-\gamma_L-\gamma_R} \ket{n,0} + \sqrt{\gamma_L} |n-1,1\rangle + \sqrt{\gamma_R} |n+1,-1\rangle$ for $n \neq 0$. (At the boundary we impose the special rule $\ket{0,0} \mapsto \sqrt{1-\gamma_0} |0,0\rangle + \sqrt{\gamma_0} |1,-1\rangle$.) 

We initialize the emitter in the state $\ket{n = 0}$, and after generating $N$ qutrits we measure it and post-select on the outcome $\ket{n = 0}$. Before the final measurement, the global quantum state can be written schematically as 
\beq
|\Psi\rangle = \sum\nolimits_{\text{histories }p} w_p |\text{end state}\rangle \otimes |\text{history } p\rangle.
\eeq
where the sum runs over all possible histories, and $w_p$ is the weight of the path (i.e., a product of $\gamma$'s). Post-selecting on the end state $\ket{n = 0}$ gives a superposition over all legal paths consistent with that end state. These paths are precisely Motzkin walks (i.e., paths that never cross the origin). Since the constraint that paths cannot cross the origin is built into the emitter dynamics, and the output state of the reservoir qubit records the time derivative of $n$, the reservoir state $\langle n = 0 | \Psi\rangle$ consists of a superposition over all Motzkin walks that start and end at $n = 0$. This is precisely the spin-1 Motzkin state explored in Refs.~\cite{PhysRevLett.109.207202, zhang2017novel}. As a function of $\delta \equiv \gamma_R/\gamma_L - 1$ the reservoir state has three regimes: when $\delta < 0$ the emitter is confined within a distance $\xi \sim |\delta|^{-1/2}$ of the origin. Therefore it can be well approximated by an MPS of bond dimension $\xi$, and the entanglement of contiguous segments of reservoir qubits scales as $\log \xi$. 

When $\delta = 0$, the emitter's state undergoes a random walk and spreads out over a distance $\sim \sqrt{N}$, before being refocused by the boundary condition at step $N$. In this case, the channel is gapless, so in principle one should not replace the evolution under the dual channel with the identity. However, if we are considering the entanglement of $\ell \ll N$ qubits, the time-evolved operator $\ket{n = 0} \bra{n = 0}$ is spread out uniformly over a much wider spatial range than the steady state, so the construction in Fig.~\ref{fig1}(c) is asymptotically exact. 
Therefore, the entanglement of this subregion of $\ell$ qubits scales as $\log \ell$. Post-selection succeeds at a rate $\sim 1/\sqrt{N}$---but this overhead is only polynomial in $N$.
Finally, when $\gamma_L < \gamma_R$, the post-selected state remains critical, but the probability of successfully post-selecting onto the $\ket{n = 0}$ state is suppressed exponentially in $N$. We will not consider this regime further.

\begin{figure}[t]
\begin{center}
\includegraphics[width=0.45\textwidth]{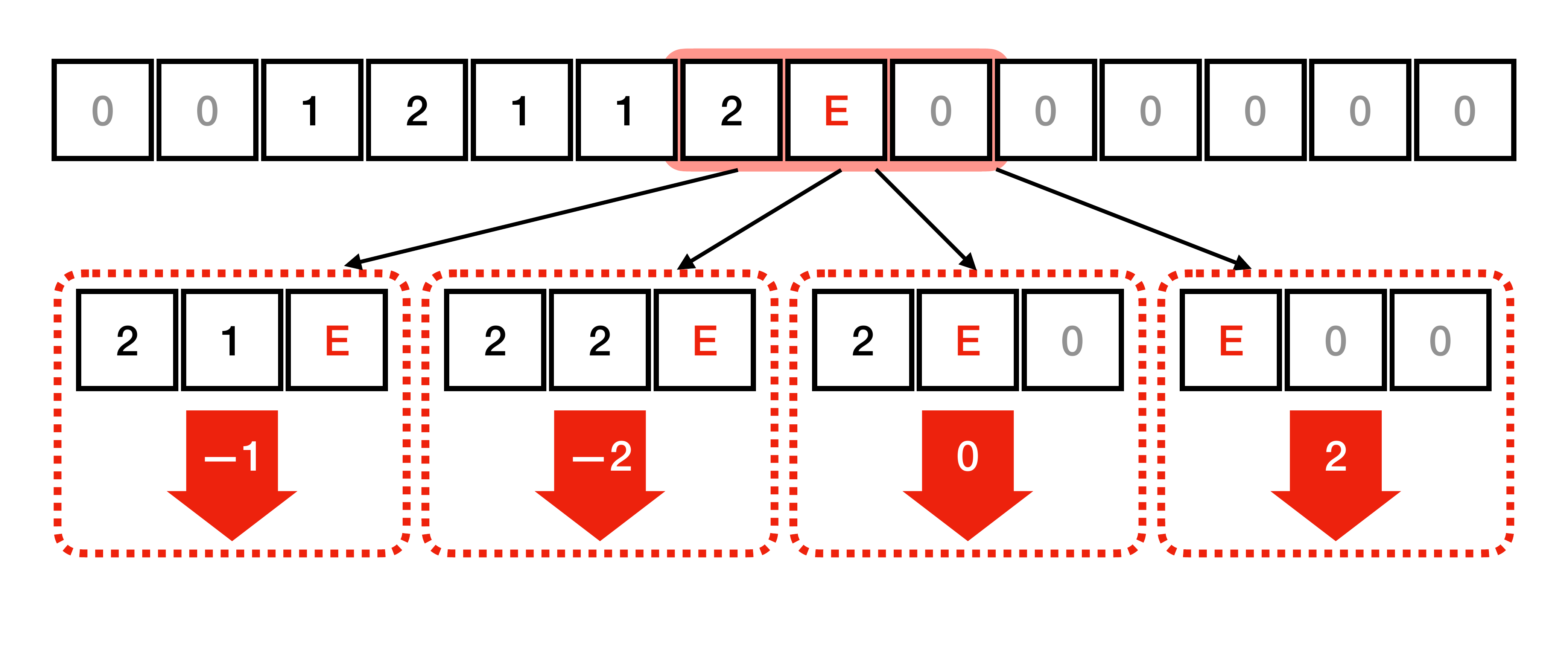}
\caption{Typical configuration of the stack for a spin-$2$ Motzkin chain. Nontrivial symbols appear between the fixed left boundary and the moving right boundary indexed by the sole $E$ site. Outside this region, the stack is in state $0$. The legal moves from this configuration are two-site moves on the stack, enumerated in the dashed red boxes. Each transition leaves a distinct imprint on the environment (or equivalently corresponds to a distinct input symbol).}
\label{fig3}
\end{center}
\end{figure}

\emph{Higher-spin Motzkin chains}.---For the spin-$1$ Motzkin chain, the post-selection overhead of the process we described yields a computational complexity that is similar to what is achievable with MPS techniques, since the bond dimension required for the MPS scales as $N$. We now turn to higher-spin Motzkin chains, in which the bond dimension of the emitter grows super-polynomially in $N$, but sequential generation remains efficient. These higher-spin chains can be realized through a simple extension of our approach above. We now need the emitter to be a many-body spin chain with four states per site, which we denote $0,1,2,E$. 
In our construction the dynamics of the spin chain is ``facilitated'' by sites in state $E$: nontrivial evolution only happens in the vicinity of such sites. 
Moreover, we assume that the number of $E$ sites is conserved, and for concreteness consider states with a single $E$ site. 
This $E$ site undergoes a random walk, and as it does so it leaves behind a nondynamical trail of $1$'s and $2$'s. Let us begin, again, by specifying the dynamics from the standard perspective. This dynamics is again completely incoherent. At each step, the neighborhood around an $E$ site can undergo the following moves: (i)~the $E$ site can move one step to the right, i.e., $E0 \mapsto 1E, 2E$, (ii)~it can move one step to the left, erasing $1$ or $2$ (but not $0$) on that site, so that $1E, 2E \mapsto E0$, or (iii)~it can remain in place (Fig.~\ref{fig3}). The initial state of the spin chain has the form $\ldots 000 E 000 \ldots$, where the initial location of $E$ is designated as the origin. Note that according to the rules above, $0E0 \mapsto 0E0, 01E, 02E$ are the only allowed moves when the $E$ site is at the origin. Therefore, depending on the rates of the allowed processes, the $E$ site undergoes a biased random walk with a reflecting boundary at the origin. Note also that to have the random walk of the $E$ site be unbiased, the total probability of moving right must match that of moving left. So the rates of the two type~(ii) processes must add up to that of the single type~(i) process. 

To construct the dual perspective, since there are five legal moves, we choose the emitted qudits to have on-site dimension $5$, and denote their states as $-2, -1, 0, 1, 2$. The unitary $U$ that accomplishes this acts on the reservoir qudit and a three-site neighborhood in the emitter, so it is spatially local. It is specified by the matrix elements (specifying only the state around the $E$ site, see Fig.~\ref{fig3}):
\bea
 \langle jkE \otimes n -_m k | U | jE0 \otimes n   \rangle & = & \sqrt{w^+} \nonumber \\
\langle 0kE \otimes n -_m k | U | 0E0 \otimes n \rangle & = & \sqrt{w'_+} \nonumber \\
\langle jE0 \otimes n +_m k | U | jkE \otimes n   \rangle & = & \sqrt{w^-} \nonumber\\ 
 \langle kE0 \otimes n | U | kE0 \otimes n\rangle & = & \sqrt{w_0}\label{2motz}.
\eea
where $+_m$ denotes modular addition on $\{ -2, -1, 0, 1, 2 \}$ and $j,k \neq 0$. When the reservoir qudit is initialized in $\ket{0}$ this unitary generates the channel described above. At the point where the random walk is unbiased, $2 w^+ = w^-$, $w'_+ = 2 w^+ + w^-$, $w'_+ + w_0 = 1$. 
%

The phase diagram of this process is very similar to that we considered for the spin-$1$ Motzkin chain, since the random-walk dynamics is essentially identical, except for each path being decorated by a trail of $1$'s and $2$'s. However, the entanglement structure of the three phases---which follows from the entropy of the emitter's density matrix at the cut---scales very differently. In the confined phase, it is still an area law, though now the bond dimension grows \emph{exponentially} with the number of sites that the walker spans in steady state. At the critical point, the entanglement of a region of size $\ell$ grows as $\sqrt{\ell}$, and this state is the spin-$2$ Motzkin state~\cite{movassagh2016supercritical, PhysRevB.94.155140}. The large bipartite entanglement of this state stems from the fact that the \emph{sequence} of unpaired $1$'s and $2$'s left of the cut is mirrored by the sequence of unpaired $-1$'s and $-2$'s right of the cut. Despite this vastly increased entanglement, the post-selection overhead for an $N$-qubit state is still just $\sqrt{N}$. Finally, in the outward-drifting phase, post-selection becomes exponentially hard. 

The procedure above gives a time complexity $\sim N^{3/2}$ (and a total number of gates $\sim N^2$) for preparing a critical Motzkin state of $N$ qudits. 
All the gates can be made manifestly spatially local, as discussed in~\cite{suppmat}. 
By contrast, the inverse gap of the Motzkin state is bounded below as $N^2$~\cite{movassagh2016supercritical} and is likely parametrically larger~\cite{PhysRevB.96.180402, PhysRevB.96.180404}. So the outlined protocol appears to give a large speedup over adiabatic state preparation.

\emph{Push-down automata}.---%
The rules we constructed for the emitter can be understood as follows: this system is a tape that stores a memory, and the $E$ site is the head of the tape. Seen this way, the emitter is a familiar object from theoretical computer science, namely a push-down automaton (PDA)~\cite{sipser2012introduction}. 
It is well known that MPSs correspond to finite-state machines~\cite{PhysRevA.78.012356}. A finite-state machine has $\chi$ internal states, of which one is the ``start'' state and one or more are ``accept'' states. Each time it receives an input, the machine undergoes a transition among its internal states, depending on the input via a transition function. After reading the entire input string, the machine is either in an accept state (so the string is ``accepted'') or not (the string is rejected). The set of all strings recognized by the machine is called the \emph{language} recognized by that machine. The correspondence with channels is as follows: the transition functions corresponding to the machine input $i$ are the Kraus operators corresponding to the measurement outcome $i$ on the reservoir qubit. Instead of generating a binary answer, each input generates a complex weight (i.e., the amplitude of that configuration in the MPS), so an MPS is a \emph{weighted} finite-state machine~\cite{PhysRevA.78.012356}.

PDAs differ from finite-state machines by containing infinite ``stacks'' where information can be stored. At each step, beyond changing its internal state, the PDA can either ``push'' information onto the top of the stack or ``pop'' information off the top of the stack. This last-in, first-out nature of the memory is the defining feature of PDAs. Evidently the machine we devised for generating Motzkin states has this form: it stores information (a string of $1$'s and $2$'s) in a stack and can only read or modify the symbol at the top of the stack. 
So we can think of it as a canonical-form quantum PDA (QPDA). We now generalize this scheme to one that prepares superpositions of strings recognized by generic PDAs. 

This construction is detailed in~\cite{suppmat}; here we discuss it schematically. The QPDA evolves according to two types of moves: (i)~an isometry from the site at the top of the stack to the neighboring empty sites, which causes the stack to grow, and (ii)~an amplitude-damping channel that resets the top of the stack to the trivial state $0$. In the Motzkin example, pushing corresponds to (i) followed immediately by (ii), while popping is just a move of type (ii). The isometry in this case is trivial as there are no correlations on the stack (but introducing correlations is straightforward, and just involves conditioning the weights in Eq.~\eqref{2motz} on the state at the top of the stack). Ensuring that this can be done through translation-invariant local unitaries is involved, but can be done through a conveyor-belt construction~\cite{suppmat} on a three-leg ladder. The uppermost leg keeps track of where the top of the stack is, and contains a single particle with position marked $E$. The middle leg holds the stack, which has an on-site Hilbert space containing symbols that represent each terminal and variable, as well as the special symbols $0$ and $X$. The lower leg is an open system with a conveyor-belt structure: 
particles are injected at the left end, move at a constant speed along the lower leg, and are extracted at the right end.

\emph{Achievable PDAs and generalizations}.---The bottleneck to preparing states \emph{efficiently} by this method is post-selection on an empty stack. In the Motzkin example, the top of the stack undergoes a one-dimensional random walk, and returns to the origin infinitely many times. More generally, if the first $n$ characters of a typical legal string of length $n' \gg n$ form a legal string with probability that scales polynomially in $1/n$, the post-selection overhead is polynomial. Many textbook examples of CFLs have this property, e.g., superpositions of strings with balanced numbers of $0$'s and $1$'s, or a cat-like state on qutrits, consisting of all strings with either equal numbers of $0$'s and $1$'s or equal numbers of $0$'s and $2$'s~\cite{sipser2012introduction}. Of course, if we relax the requirement of time-translation invariance then many more PDAs become achievable, including the volume-law regime of the spin-2 Motzkin state. This state can be constructed by biasing the stack to push particles on until $N/2$ particles are on the stack, then reversing the bias. We have not focused on these cases as they give states with simple entanglement patterns consisting of distant Bell pairs. 

The perspective discussed here generalizes to machines that are not, strictly speaking, PDAs: for example, the channel could have transition rates that are random functions of the position of the head of the stack. If we allow for random rates, the random walk of the $E$ particle can become subdiffusive instead of diffusive. This allows one to realize critical Motzkin-like states for which the entanglement of subsystems of size $\ell$ scales as $S(\ell) \sim \ell^{\alpha}$ for $\alpha \in (0, 1/2)$. The $E$ particle can also undergo superdiffusive L\'evy walks~\cite{bouchaud1990anomalous}, by the following construction. Instead of having the bias of the random walk be fixed, we promote it to a dynamical variable, controlled by the the instantaneous state of a dynamical switch~\cite{suppmat}. This switch, in turn, is controlled by an auxiliary quantum walker: whenever this walker returns to its origin, the bias of the Motzkin walk flips. Suppose this quantum walker is diffusive in one dimension~\footnote{While this is not the most natural situation it is achievable using the Aubry-Andr\'e model at its critical point, or the random dimer model~\cite{PhysRevLett.65.88}.}: after a time $t$ it switches the flip at a rate $t^{-1/2}$, so the mean free path (and therefore the diffusion constant) scales as $t^{1/2}$. Therefore the displacement scales as $x \sim t^{3/4}$ and by the duality we have discussed the Motzkin state has bipartite entanglement $S(\ell) \sim \ell^{3/4}$. To ensure a pure state we post-select at the end on the quantum walker returning to its origin, leading to an extra $N^{1/2}$ overhead for a chain of size $N$. A similar construction using a ballistic quantum walker in one dimension or a diffusive walker in two dimensions gives entanglement scaling as $S(\ell) \sim \ell / O(\log \ell)$, with additional post-selection overhead $\sim N$ (i.e., still polynomial). 

\emph{Discussion}.---We have explored a recipe for sequentially preparing quantum states; this recipe consists of harvesting the radiation from an open quantum system, and post-selecting on the final state of the open system to arrive at a pure state on the radiation. When the open system undergoes a phase transition, so does the radiated state. These observations lead to efficient state-preparation schemes when the post-selection overhead is controllable, i.e., when the entropy of the open-system density matrix stays small. We have described one class of systems in which this can happen: namely, canonical QPDAs, in which the entropy of the density matrix is controlled by the motion of the head of a stack. Other classical stochastic models with low-entropy steady states occur, e.g., at absorbing-state transitions~\cite{cardy1980directed}; exploring these is an interesting direction for future work. Even when the protocol we have outlined is inefficient, the idea of reverse-engineering quantum states from stochastic processes might prove useful as a means of discovering exotic states in higher-dimensional quantum systems~\cite{zhang2022coupled}, or computing their entanglement or correlation properties (since the corresponding channels are classical Markov chains~\footnote{W. Zhang, F. Pollmann, and S. Gopalakrishnan, in preparation.}). 
The emergence of PDAs in the quantum context also motivates further study of the statistical mechanics of weighted PDAs. 
At present many questions about these systems remain unsettled, 
 including the conditions under which states generated by PDAs are ground states of frustration-free local Hamiltonians~\footnote{This question will be addressed in detail in a forthcoming paper by Balasubramanian, Lake, and Choi.}. 

Finally, while the constructions here are perhaps too involved to be directly relevant to near-term experiments, the conveyor-belt construction is related to experiments in quantum optics in which a many-body system interacts with, e.g., photons in a waveguide. It would be interesting to investigate experimentally feasible protocols for sequentially generating highly entangled states in that context.

\emph{Acknowledgments}.---I am grateful to Shankar Balasubramanian, Soonwon Choi, Juan P. Garrahan, David Huse, Matteo Ippoliti, Vedika Khemani, Ethan Lake, Frank Pollmann, Tibor Rakovszky, Hans Singh, Romain Vasseur, and Wucheng Zhang for stimulating discussions and feedback. This work was supported by NSF DMR-1653271.

\bibliography{qpdabib}

\begin{widetext}

\section*{Details of the spin-2 Motzkin construction}

We now present a more detailed construction of the process that generates the spin-2 Motzkin state, making the role of locality manifest. This ``two-leg'' construction is a simplified version of the more general construction we will describe below. The upper leg contains the stack, as described in the main text, with its head marked by the symbol $E$, and with only $0$'s to the right of the $E$ symbol. The lower leg contains particles that move ballistically, e.g., via a swap circuit. The unmarked sites on the lower leg are in a completely inert state that does not participate in any nontrivial gates. There are two dynamical processes, which alternate in time. The first is a circuit acting on triangular four-site neighborhoods of the sort shaded in Fig.~\ref{app0}. These gates are applied in a three-layer sequence such that exactly one gate acts on each triangle. The second process is a swap circuit that moves all particles on the lower leg one step to the right. This can be implemented as a two-layer brickwork circuit since we control when the initial states are fed in. When particles reach the rightmost position on the stack (which we can take to be any position $\agt \sqrt{N}$) they are swapped out to Bob's system. Bob discards the unmarked particles as they contain no information about the dynamics. The state is defined on the remaining particles.

We now describe the structure of the four-site gate acting on a downward-pointing triangle. In principle, this gate acts on a rather large Hilbert space. We will use notation $\ket{xyz,a}$ to denote the states on the three upper sites (arranged left to right) and finally the state at the bottom of the triangle. In principle this gate acts on a rather large Hilbert space, but it acts as the identity on most configurations. The gate acts nontrivially only on the subspace containing the following six states: 
\begin{equation}\label{processlist}
\ket{1E0,0}, \ket{2E0,0}, \ket{E00,1}, \ket{E00,2}, \ket{11E,-1}, \ket{12E,-2}.
\end{equation}
The weights in this subspace are specified in the main text.

\begin{figure}[b]
\begin{center}
\includegraphics[width=.75\textwidth]{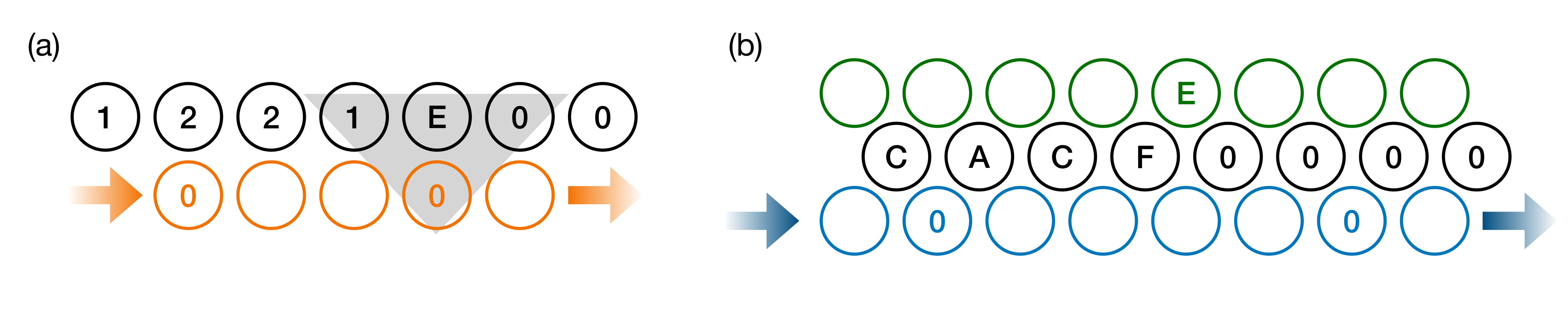}
\caption{Illustrations of typical configurations of the system in the simplified two-leg case for generating Motzkin states (a) and the general case for CFLs in Chomsky normal form~(b).}
\label{app0}
\end{center}
\end{figure}

The second process, meanwhile, moves the symbols along the lower leg at each time step. We now argue that this continuous advection ensures that the dissipative process remains Markovian: each live site on the lower leg is involved in exactly one nontrivial gate. 

The qudit is injected in state $\ket{0}$, and the first time a nontrivial gate can act is when the $E$ is situated above the $0$ (see Eq.~\eqref{processlist}). At this point, three things can happen on a particular branch of the process.

\begin{enumerate}
\item The gate acts as $\ket{xE0,0} \to \ket{E00,x}$ for $x = 1,2$. After that the lower leg moves one step right, so the state on the lower leg is now two steps ahead of $E$. The circuit geometry ensures that the $E$ will never catch up, so no more nontrivial gates act on the qubit on the lower leg.
\item The gate acts as $\ket{xE0,0} \to \ket{xE0,0}$. Then the lower leg is advected one step. At the next step, the potentially nontrivial triangle on which the gate acts is initially in the configuration $\ket{E00,0}$, which is not on the list~\eqref{processlist}. So the $E$ does not move at this step, but the lower-leg particle does. Once again, the lower leg particle is now two steps ahead of the $E$ and the $E$ cannot catch up.
\item The gate acts as $\ket{xE0,0} \to \ket{xyE,-y}$ for $x,y = 1,2$. After the lower leg moves one step right, the next two iterations of the triangle see initial states $\ket{yE0, -y}$ and $\ket{E00,-y}$, which are not on the list~\eqref{processlist}. After these steps the $E$ can no longer catch up with the lower leg. 
\end{enumerate}

This process, as we have outlined it, consists entirely of local gates. We estimate the number of gates and layers needed for the critical Motzkin state. At each timestep, we implement $O(\sqrt{N})$ gates at $O(1)$ circuit depth. We must run the circuit for $O(N)$ timesteps to collect $N$ qudits on the lower leg. Finally there is the $O(N^{1/2})$ post-selection overhead. These combine to give the estimates in the main text.

\section*{General construction of QPDAs}

A PDA recognizes the set of strings belonging to a context-free language (CFL)~\cite{sipser2012introduction}. A CFL has an alphabet consisting of variable and terminal symbols (denoted by upper and lower case letters respectively). Each terminal is inert, and each variable can evolve according to one or more substitution rules. In Chomsky normal form~\cite{sipser2012introduction} all substitution rules can be written as $A \to BC$ (i.e., a variable splits in two) or $A \to a$ (i.e., a variable is substituted by a terminal). A variable can have many different substitution rules. To construct a legal string in the CFL, one begins with the special ``start'' variable $S$, and keeps applying substitution rules until one ends up with a string consisting entirely of terminals. The variable $S$ is special in not being allowed to appear on the right-hand side of any substitution rules. 

To construct an equivalent CQPDA, we proceed as follows. As discussed in the main text the emitter in the fully general case is a three-leg ladder.  When one of the injected particles reaches the head of the stack, the symbol at the head of the stack (which in our construction will always be a variable) undergoes a superposition of all substitution rules. On branches where the variable splits into two, the injected particle moves along the stack, reaches the head again, and the process repeats. On branches where the variable is replaced by a terminal, the state of the terminal is written on the injected particle and erased from the stack. This process repeats until enough qubits have been harvested; after post-selecting on an empty stack, the harvested qubits are in a weighted superposition of all strings in the CFL. 

The ``particles'' (i.e., states that interact nontrivially with the rest of the system) are initially in state $0$. The unmarked sites on the top and bottom legs can be regarded as being in some state $x$ that does not engage in nontrivial interactions. 

\begin{figure}[b]
\begin{center}
\includegraphics[width=0.95\textwidth]{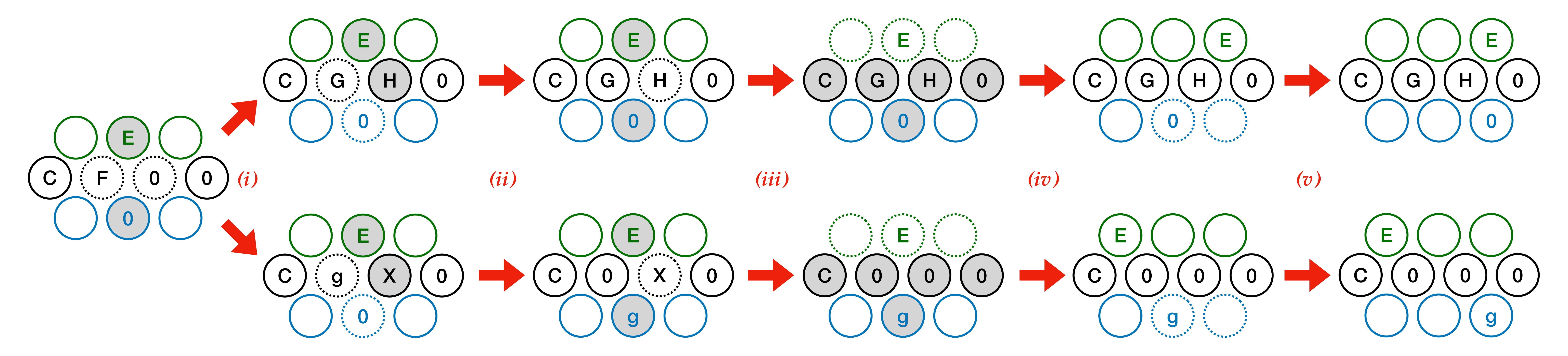}
\caption{Single cycle of gates acting on a general CQPDA, and their action in the case of an update rule that replaces a variable with two variables (upper) and in the case of an update rule that replaces a variable with a terminal (lower). The sites that are controlling the unitary to be applied at the next step are shaded in gray, and the sites that the unitary will rearrange are marked by dotted lines.}
\label{appfig}
\end{center}
\end{figure}

At the beginning of each step, the configuration of the circuit near the top of the stack is as shown in Fig.~\ref{appfig}: the $E$ site marks the boundary between the occupied and unoccupied parts of the stack, and the top of the stack is a variable. Unless a particle on the lower leg is immediately beneath the $E$ site at the start of the cycle, nothing happens during the cycle, except that the $E$ particle moves to the right. This last stage in the cycle can be implemented, e.g., by a swap circuit, assuming the $0$ particles are placed correctly. During each step, the gate involved in that step is applied in a brickwork pattern throughout the chain. This is manifestly necessary since different branches of the state have the $E$ site in different places.

If the cycle begins with an $E$ right above a $0$, then in the first step of the cycle the variable at the top of the stack undergoes a superposition of all legal moves. Because we assumed Chomsky normal form, each branch either replaces the variable with two variables or with one terminal. These cases are considered separately in the lower panel of Fig.~\ref{appfig}. For reasons that will soon be apparent, for the rules that use a terminal we insert an $X$ symbol just right of the terminal. In step 2, we perform a gate controlled on the $E$ site and the site to its southeast: conditional on the latter site being an $X$, we swap the two dotted sites. This rule ensures that terminals (and only terminals) will be popped from the stack and emitted. After this step, we apply a gate controlled on $E$ and a nontrivial symbol directly beneath it to flip the $X$ back to a $0$. After these steps, the stack has evolved, but the $E$ site is no longer at the head of the stack. We restore it by moving the head of the stack in a direction conditional on the state of the bottom leg. At this point the cycle is essentially complete; the conveyor belt on the bottom leg of the ladder moves one site and the process repeats. 

The key feature of this process that allows it to represent the CQPDA dynamics is that when a variable splits into multiple variables (so there is nothing to emit), The lower-leg particle continues to interact with the stack, since it and the head of the stack move right at the same rate. On the other hand, when a terminal is generated, the terminal is immediately emitted into the lower leg and at the end of the cycle this emitted particle has ``outrun'' the stack and will not interact with it further. This allows the device to allow for arbitrarily many moves of the stack until a terminal is generated. 

\section*{Adding switches}

It is straightforward to generalize the conveyor-belt construction above to add switches. A switch can sit to the left of the first site on the stack. The state of the switch is imprinted on particles as they enter the conveyor belt. The transition rule (i) in Fig.~\ref{appfig} is controlled by the state of the particle on the conveyor belt, which is determined by the state of the switch at the time the particle passed the switch. Therefore the weights involved in this transition rule can be controlled by the switch. As discussed in the main text, the dynamics of the switch itself can be controlled by an auxiliary system.

\end{widetext}

\end{document}